# The Impact of Genomic Variation on Function (IGVF) Consortium

Authors: IGVF Consortium (See below for detailed author list).
Correspondence to: igvf-markerpapercorrespondence@gowustl.onmicrosoft.com

## Abstract

Our genomes influence nearly every aspect of human biology from molecular and cellular functions to phenotypes in health and disease. Human genetics studies have now associated hundreds of thousands of differences in our DNA sequence ("genomic variation") with disease risk and other phenotypes, many of which could reveal novel mechanisms of human biology and uncover the basis of genetic predispositions to diseases, thereby guiding the development of new diagnostics and therapeutics. Yet, understanding how genomic variation alters genome function to influence phenotype has proven challenging. To unlock these insights, we need a systematic and comprehensive catalog of genome function and the molecular and cellular effects of genomic variants. Toward this goal, the Impact of Genomic Variation on Function (IGVF) Consortium will combine approaches in single-cell mapping, genomic perturbations, and predictive modeling to investigate the relationships among genomic variation, genome function, and phenotypes. Through systematic comparisons and benchmarking of experimental and computational methods, we aim to create maps across hundreds of cell types and states describing how coding variants alter protein activity, how noncoding variants change the regulation of gene expression, and how both coding and noncoding variants may connect through gene regulatory and protein interaction networks. These experimental data, computational predictions, and accompanying standards and pipelines will be integrated into an open resource that will catalyze community efforts to explore genome function and the impact of genetic variation on human biology and disease across populations.

## Introduction

Since the initial sequencing of the human genome, genetic studies have been immensely productive in identifying genomic variants and associating those variants with phenotypes[1–3]. Exome and genome sequencing studies have already observed hundreds of millions of genomic variants, including single-nucleotide variants (SNVs), small insertions and deletions (indels), and larger structural variants (**Fig. 1**)[4,5]. Comparisons within families, case-control cohorts, and population-scale biobanks have now identified hundreds of thousands of associations between such variants and phenotypes in both health and disease[6–12].

The next challenge is to understand how genomic variation affects molecular and cellular processes ("genome function") to influence organismal phenotype (**Fig. 1**). At a molecular level, genomic variation can impact the expression, activity, or localization of genes and proteins. Altered gene expression or protein activity can, in turn, impact the activity of other genes and proteins via networks of physical or functional interactions. Changes in molecular networks can then influence the behavior of cells and tissues, and in doing so can influence organismal phenotypes. We note that we use the term "genome function" to refer to these molecular and cellular processes encoded by the genome, and note that this does not necessarily imply "function" in terms of organismal or evolutionary selection.[13,14]

Previous and ongoing efforts have produced breakthroughs in mapping various aspects of genome function, including locating and annotating millions of noncoding regulatory elements in the human genome[15,16]; mapping associations between genomic variants and effects on gene or protein expression across dozens of human tissues[17,18]; profiling hundreds of cell types and states through single-cell measurements of gene expression[19,20]; applying saturation mutagenesis to analyze coding variants in selected disease genes[21–23]; and characterizing how genes and proteins interact genetically or physically in molecular networks[24–26]. These and other studies have also demonstrated how mapping the impacts of genomic variation on genome function can reveal molecular mechanisms in human biology and disease, guide genetic diagnosis and clinical management, and facilitate the development of novel therapeutics (**Fig. 1**, reviewed in[1,27,28]). In instances when disease mechanisms have been revealed, there are often accompanying advances in understanding basic biology with far-reaching benefits beyond the specific disease of study.

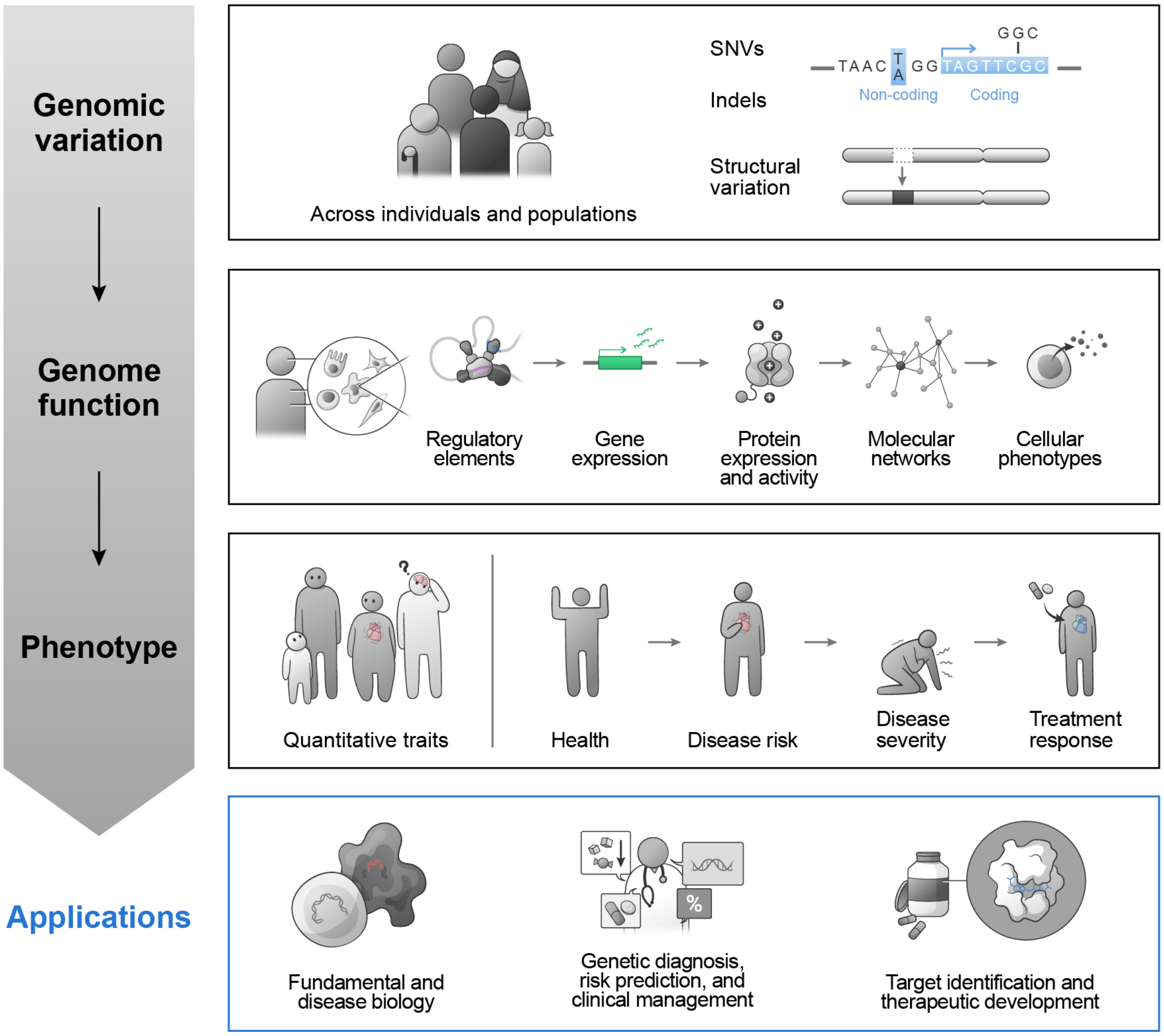

**Figure 1. Genomic variation influences genome function and phenotype.**

Yet, connecting genomic variants to functions and phenotypes continues to prove challenging, and numerous obstacles have blocked rapid progress. The sheer number of genomic variants, both those we have observed already and those we might observe in the future, is immense, and we lack any perturbation-based data for most variants. Due to linkage disequilibrium, most genetic associations with common diseases contain many candidate variants, and the variant(s) that causally affect disease risk are unknown. The vast space of possible molecular and cellular effects has been challenging to study systematically: for example, coding variants might affect protein function via effects on stability, localization, or interactions with proteins; noncoding variants might affect gene expression through effects on transcription factor binding, chromatin state, and regulatory interactions; and genes and proteins might impact cellular processes through diverse mechanisms involving gene regulatory networks, signaling pathways, protein complexes, and other interactions. Genomic variants, elements, and networks can also have highly cell-type or context-dependent activities, yielding additional complexity given the large number of cell types in the human body. Finally, while previous efforts have largely focused on individual layers of genome function, such as studying coding variants or annotating noncoding regulatory elements, understanding the impact of genomic variation on phenotypes and disease may require a more holistic, integrative understanding of genome function that connects molecular to cellular to physiological processes. Due to these and other challenges, the molecular mechanisms underlying many genetic associations for common diseases remain to be established[2,29], and genetic diagnosis for rare diseases continues to be hindered by the preponderance of variants of uncertain significance (VUS)[7,30]. New coordinated research activities will be needed to address the scale and scope of these challenges and thereby unlock the vast unrealized potential for understanding human biology and for improving human health[31,32] .

With these challenges in mind, the National Human Genome Research Institute (NHGRI) launched the IGVF Consortium in 2021, with the goal of developing a systematic understanding of the effects of genomic variation on genome function and how these effects shape phenotypes. The Consortium consists of >120 laboratories collaborating on five key activities: (i) Mapping Centers, to analyze regulatory element and gene activity at single-cell resolution across hundreds of cell types; (ii) Functional Characterization Centers, to systematically characterize the molecular and cellular effects of introducing variants or perturbing elements and genes; (iii) Predictive Modeling Projects, to develop and apply computational approaches to comprehensively model the impact of genomic variation on genome function and guide experimental design; (iv) Regulatory Network Projects, to advance network-level understanding of the influence of genetic variation and genome function on cellular and organismal phenotypes; and (v) Data and Administrative Coordinating Centers, to lead development of resources and infrastructure to share IGVF data, standards, and pipelines with the scientific community. IGVF membership and activities are expanding further via Affiliate Membership, a process by which any researcher or research project can apply to join IGVF to drive its vision and execution. Through these activities, the IGVF Consortium aims to generate an extensive resource of experimental data, standardized protocols, and computational tools integrated into a catalog that can be broadly deployed for exploring genome function and the impact of genetic variation on human biology and diseases in diverse populations. Below we describe the goals, strategies, and anticipated deliverables of IGVF (**Box 1**).

**Box 1: IGVF goals and approaches**

- Characterize the impact of genomic variants, regulatory elements, and genes on molecular and cellular phenotypes — by analyzing millions of naturally occurring or designed genomic perturbations across dozens of cellular models.
- Identify where and when regulatory elements and genes are active with resolution for individual cell types and states — by applying single-cell mapping technologies across hundreds of biological samples including cellular models, tissues, and environmental contexts.
- Predict the consequences of genomic variation on genome function and phenotype for previously unstudied variants and/or cellular contexts — by developing predictive computational models that can generalize across contexts and establishing benchmarking pipelines to evaluate and calibrate their accuracy.

- Study diverse cellular and disease systems, types of genomic variation, and aspects of genome function — by developing and applying a "map-perturb-predict" framework in which single-cell mapping, genomic perturbations, and predictive modeling are synergistically combined.
- Create an initial map that annotates the predicted effects of every possible single-nucleotide variant in the human genome on key aspects of genome function — by integrating models for how coding variants might alter protein function, how noncoding variants might affect gene expression, and how noncoding and coding variants might connect within molecular networks.
- Advance our understanding of the impact of genomic variation on disease — by exploring how best to apply IGVF resources to inform genetic diagnosis and to identify biological mechanisms of disease risk.
- Ensure that these advances are applicable to and inclusive of people of diverse sexes, ancestries, and populations — by studying individuals with different genetic backgrounds, assaying and predicting effects of variants observed in diverse populations, and studying diseases disproportionately affecting disadvantaged or under-represented populations.
- Catalyze research by others toward the long-term goal of understanding the impact of genomic variation — by partnering with the broader research community and developing resources and infrastructure to share IGVF data, methods, standards, and pipelines.

## Connecting genomic variation to effects on genome function and phenotype via a map-perturb-predict framework

To create a comprehensive catalog of the effects of genomic variation, IGVF has developed a strategy that integrates three complementary components (**Fig. 2**). One component will be to quantify the activity of regulatory elements and the expression of genes via single-cell mapping. Another will conduct systematic perturbations of variants, regulatory elements, and genes. A third will seek to generalize results to new, unstudied genomic variants and cellular contexts via predictive modeling. By integrating these three components in a map-perturb-predict framework, we aim to achieve substantial synergy across the consortium.

### *Mapping the activities of genes and regulatory elements at single-cell resolution*

Identifying noncoding regulatory elements and genes and mapping their activities across cell types and states is foundational for understanding where and when genomic variation might impact genome function. Due to technological limitations, many previous efforts have lacked this level of resolution. Recent advances in single-cell technologies now enable the generation of comprehensive maps of chromatin state and gene expression in nearly any cell type in the body[19,20], and computational analysis of these datasets can help to locate candidate regulatory elements, correlate element and gene activities, identify transcription factor (TF) binding regions and footprints, and reveal molecular pathways[33–35]. We will collect single-cell data across hundreds of cell types and states (see below for biological systems and contexts). We will apply primarily single-nucleus (sn)ATAC-seq and snRNA-seq, including in multiomic formats, and explore new single-cell approaches for TF binding, histone modifications, chromatin interactions, and clonal tracing. These data will provide a foundation for interpreting the effects of functional characterization experiments and building cell-type-specific maps of variant effects.

*Functional characterization of variants, regulatory elements, and genes via genomic perturbations*

Perturbation experiments will be crucial for understanding the causal relationships among variants, regulatory elements, genes, and phenotypes, but until recently have been challenging to apply at sufficient scale. New enabling technologies include high-throughput genetic screens using CRISPR genetic or epigenetic perturbations and over-expression strategies[21,22,36–43]; massively parallel reporter assays (MPRAs) to quantify enhancer and promoter activities of noncoding sequences and their variants[44–50]; and studies of naturally occurring genetic variation to identify and fine-map different types of quantitative trait loci (QTLs)[17,51,52]. IGVF plans to conduct millions of experimental perturbations, including to directly study the effects of naturally occurring or designed DNA variants, and to perturb regulatory elements and genes to build maps of genome function. We will characterize the effects of these perturbations using diverse assays including measurements of chromatin accessibility[53], gene expression[54–56], protein expression and activity[24,57–60], and molecular and cellular phenotypes[61]. These data will enable directly characterizing variants of interest, such as those associated with disease, and provide data to train or evaluate predictive models of variant effects.

*Predictive models of genomic variation and genome function*

Genome function is complex, and we cannot expect to experimentally map the effects of all possible variants on all possible activities in all possible cellular contexts. To address this, recent studies have highlighted the possibility of developing powerful predictive models that can make predictions that generalize across contexts — for example, to link genetic variants to effects on TF binding and chromatin accessibility[50,62–65]; identify TF footprints[33,65]; connect regulatory elements to their target genes[65–67]; or identify causal genes and cell types enriched for heritability for complex diseases or traits[68–73]. Equally importantly, successes by CASP[74], ENCODE[15], and others[17,75] have illustrated how developing uniform standards, gold-standard datasets, and benchmarking pipelines can catalyze advances throughout the global scientific community by enabling rigorous evaluation of accuracy and direct comparison of alternative strategies. We will leverage new advances in machine learning and large-scale perturbation datasets across cell types and contexts to tackle key prediction problems — including mapping aspects of genome function, interpreting the impact of genomic variation, and guiding the design of future experimental assays such that the data produced will be maximally informative for subsequent predictive modeling. To systematically evaluate and calibrate such models, we will build benchmarking pipelines that compare predictions to perturbation data, including both from IGVF functional characterization experiments and external sources such as QTL, GWAS, and genome sequencing studies. In areas where data collection is already advanced, we will engage the external community by designing prediction challenges with held-out validation datasets produced by IGVF.

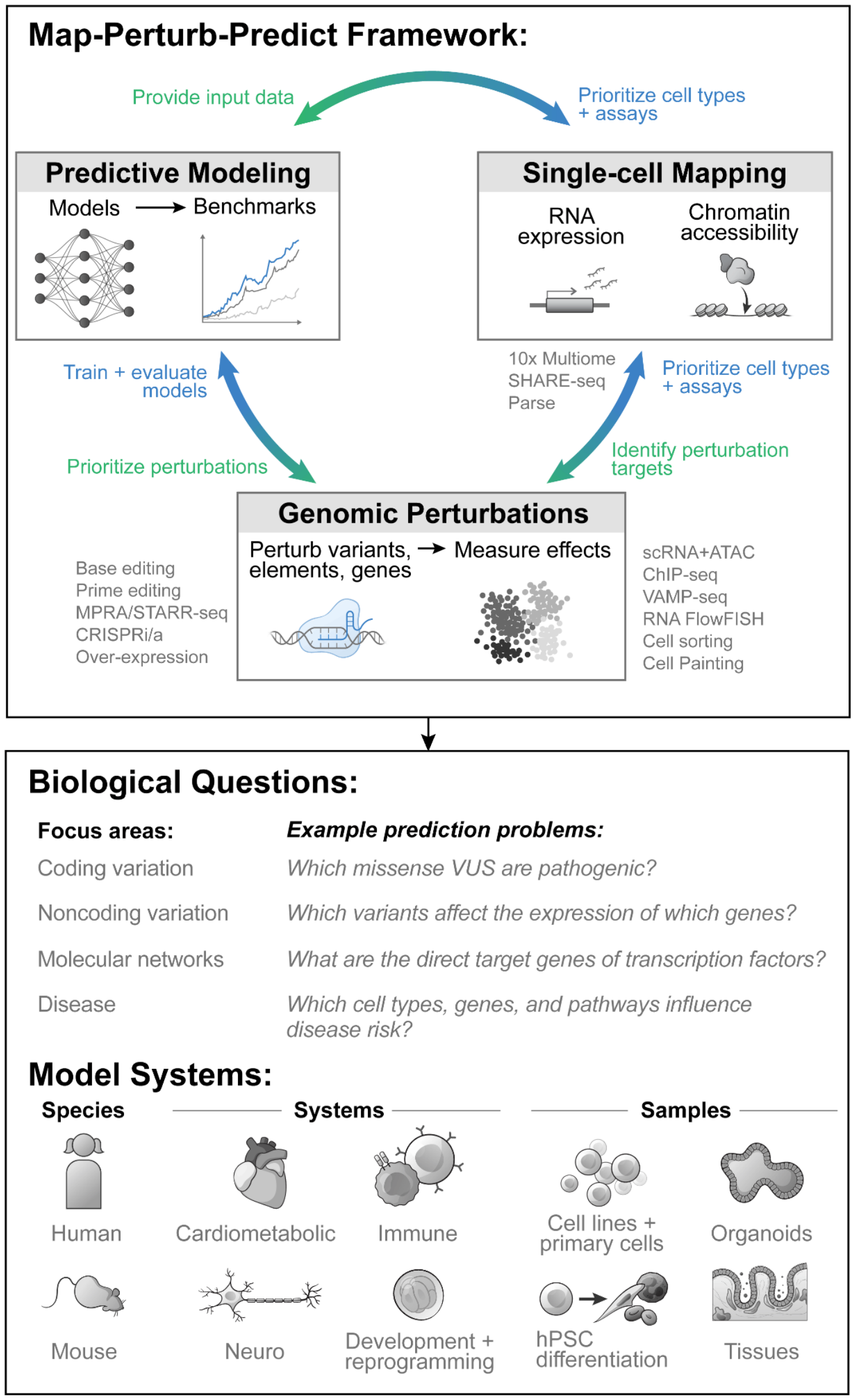

**Figure 2. A map-perturb-predict framework to connect genome variation to genome function and phenotype**

*Applying the map-perturb-predict framework to study genomic variation and genome function across cellular and biological systems*

Together, these three activities will form an iterative map-perturb-predict framework (**Fig. 2**) that we will apply to study various aspects of genomic variation, genome function, and phenotype. IGVF projects will investigate single-nucleotide variants, indels, and structural variation, and map the relationships between elements, genes, proteins, and their molecular networks in diverse cellular states and phenotypes (**Fig. 1**). IGVF projects will study a variety of biological systems, including iPSC models (2D and 3D) differentiated into lineages spanning all germ layers; primary cell types relevant to disease areas of interest, including cardiometabolic, immune, neuropsychiatric, and neurodevelopmental diseases; and tissues *in vivo* to inform how cell-cell interactions and environment alter genome function (**Fig. 2**). The selected models will include dynamic biological processes that will provide insights into how regulatory networks change over time, such as B cell activation and differentiation or fibroblast-to-iPSC reprogramming. While the primary objective of IGVF is to characterize variation and function of the human genome, IGVF studies will also create resources and leverage mouse models for certain studies, such as for *in vivo* CRISPR screens to understand how genes affect cellular phenotypes in a tissue environment, and for comparing the effects of variants, elements, and genes across individuals with different genetic backgrounds. Together, these areas of exploration will yield insights about genomic variation and genome function across diverse areas of biology and help to identify optimal strategies that can be more broadly applied to additional biological systems.

## Map of genome function and variant effects integrating coding variation, noncoding variation, and molecular networks

An integrative deliverable for IGVF will be to generate a preliminary variant-effect map that integrates three key aspects of genome function: gene expression, protein activity, and molecular networks. This draft map would allow querying, for any possible single-nucleotide variant in the genome: Is this variant measured or predicted to (i) impact transcription factor binding, regulatory element activity, and target gene expression in particular cell contexts, for noncoding variants; (ii) impact protein function, for coding variants; and (iii) connect to other genes/proteins via gene regulatory networks and/or protein-interaction networks, for both coding and noncoding variants? We will integrate this map of genome function, along with external data, into a multi-relational knowledge graph[76–78] that can be readily queried by users as part of the IGVF Catalog (see below, **Fig. 3**).

For each of these aspects of genome function, existing computational models have been shown to have some utility in understanding the impact of genomic variation on diseases and traits, but much work is needed to improve the accuracy of these models and to conduct systematic evaluations using more precise and comprehensive gold-standard datasets. To address this, we will establish a pipeline to benchmark all predictions against functional characterization datasets and external human genetics datasets — allowing us to rigorously evaluate and guide interpretation of the draft map. We anticipate that providing genome-wide predictions from the best models, together with a reproducible benchmarking framework, will help launch an iterative and ongoing effort extending beyond IGVF to improve the accuracy of this map over time (**Fig. 3**).

### *Effects of noncoding single-nucleotide variants on regulatory element activity and target gene expression*

In the 99% of our genome that does not encode for proteins, noncoding variants can impact genome function by altering gene expression or regulation. While previous studies have mapped regulatory elements, gene expression patterns, transcription factor binding, and the effects of variants on gene expression in tissues or cells, we still lack models that can make accurate causal inferences about how genomic variation affects gene regulation[79–81]. We will seek to build genome-wide annotations of key components of this *cis*-regulatory code: Which single-nucleotide variants affect transcription factor binding sites, regulatory element activity, and gene expression in *cis*, in which cell types or states, with what magnitude and direction of effect?

To do so, IGVF plans to (i) generate multiomic snRNA+ATAC-seq data to a depth needed to comprehensively identify candidate cis-regulatory elements, detect transcription factor footprints[33], and predict enhancer-gene relationships[34,35,65,67]; (ii) test >1 million noncoding variants in enhancer activity reporter assays[44,45,49,50,82]; (iii) test >100,000 noncoding variants for effects on expression through fine-mapping of eQTLs or direct CRISPR-based genome editing[17,36–38,40,51]; (iv) measure >100,000 putative regulatory interactions between candidate regulatory elements and nearby genes, for example using dCas9-based epigenome editing paired with single-cell readouts of RNA expression[66,83–86]; (v) and perturb transcription factors to read out effects on gene expression using Perturb-seq[54–56]. The variants and elements studied will include both naturally occurring and designed sequences, which will be critical for building accurate models of the gene regulatory code[87]. Each of these experiments will be conducted in multiple cellular models, so that the data can be used to refine and develop predictive models that can construct maps of noncoding variant effects across many cell types.

*Effects of single-nucleotide variants in protein-coding genes on function*

For protein-coding sequences, our ability to interpret the functions of genomic variation is based on our knowledge of the genetic code for protein synthesis — which has enabled identifying open reading frames encoding novel proteins, identifying null, frameshift or nonsense variants, and predicting damaging missense variants. However, missense variants and inframe indels remain difficult to interpret, and we still lack a comprehensive understanding of how changes in protein sequence might affect different aspects of protein structure, expression, dynamics, and activity, including the impact on stability, subcellular localization, or interactions with other proteins.

We will improve annotations of how protein-coding missense variants impact protein stability and activity by applying high-throughput technologies[24,57–60] to experimentally characterize the impacts of >200,000 missense variants on various properties of proteins and their phenotypic impacts in cellular models, including protein stability, subcellular localization, cell viability, cell morphology, and protein-protein interactions. These experiments will focus on clinically relevant genes, such as those associated with Mendelian diseases, to provide direct look-up tables for certain genes, and provide data to refine or develop new predictive models to predict the likely impact of any coding variant across the genome.

*Linking noncoding and coding variants to gene regulatory and protein interaction networks and selected cellular phenotypes*

Upon linking a variant to effects on gene expression or protein activity in *cis*, we will seek to annotate the sets of other genes and proteins linked to the variant in *trans* through molecular networks in a given cell type or state. To a more limited extent, we will explore links to downstream cellular phenotypes. Genes and proteins can work together in many different ways, and it has been challenging to map or infer these sets of functionally related genes and corresponding cellular phenotypes in a comprehensive and cell-type specific fashion.

To construct molecular networks, we will focus on defining (i) gene expression programs, described by sets of genes whose expression levels are correlated across single cells; (ii) gene regulatory networks that infer which transcription factors directly regulate which target genes via particular regulatory sequences; (iii) sets of interacting proteins or protein complexes; and (iv) dynamic changes to these programs/networks across cell state transitions.

To build these maps, we will collect longitudinal multiomic RNA and ATAC-seq data across dynamic cellular processes including differentiation and reprogramming[88–91]; study how genes and proteins interact in molecular networks, including by mapping protein-protein interactions[24] and conducting large-scale Perturb-seq[54–56]; and assessing how CRISPR-based perturbations or natural genetic variation across individuals affects cellular phenotypes including differentiation, gene expression programs, and cellular states. We will establish benchmarks to evaluate how best to use these data to construct cell-type and state-specific molecular networks and assess the impact of genomic variation on cellular phenotypes.

We anticipate that many aspects of this map of genome function and variant effects will be cell-type specific, with annotations for each of the hundreds of cell types, states, and contexts studied by IGVF. This could be accomplished by developing predictive models that use multiomic snRNA-seq and ATAC-seq as their only cell-type specific input data[34,35,67,92].

The research infrastructure IGVF develops to build these maps will set in motion community efforts to expand on this framework by collecting additional datasets, training improved models, generating more accurate maps, and expanding the approach to additional cell types and aspects of genome function. This draft map will also offer immediate opportunities to address questions about the impact of genomic variation and genome function on phenotypes (see next section).

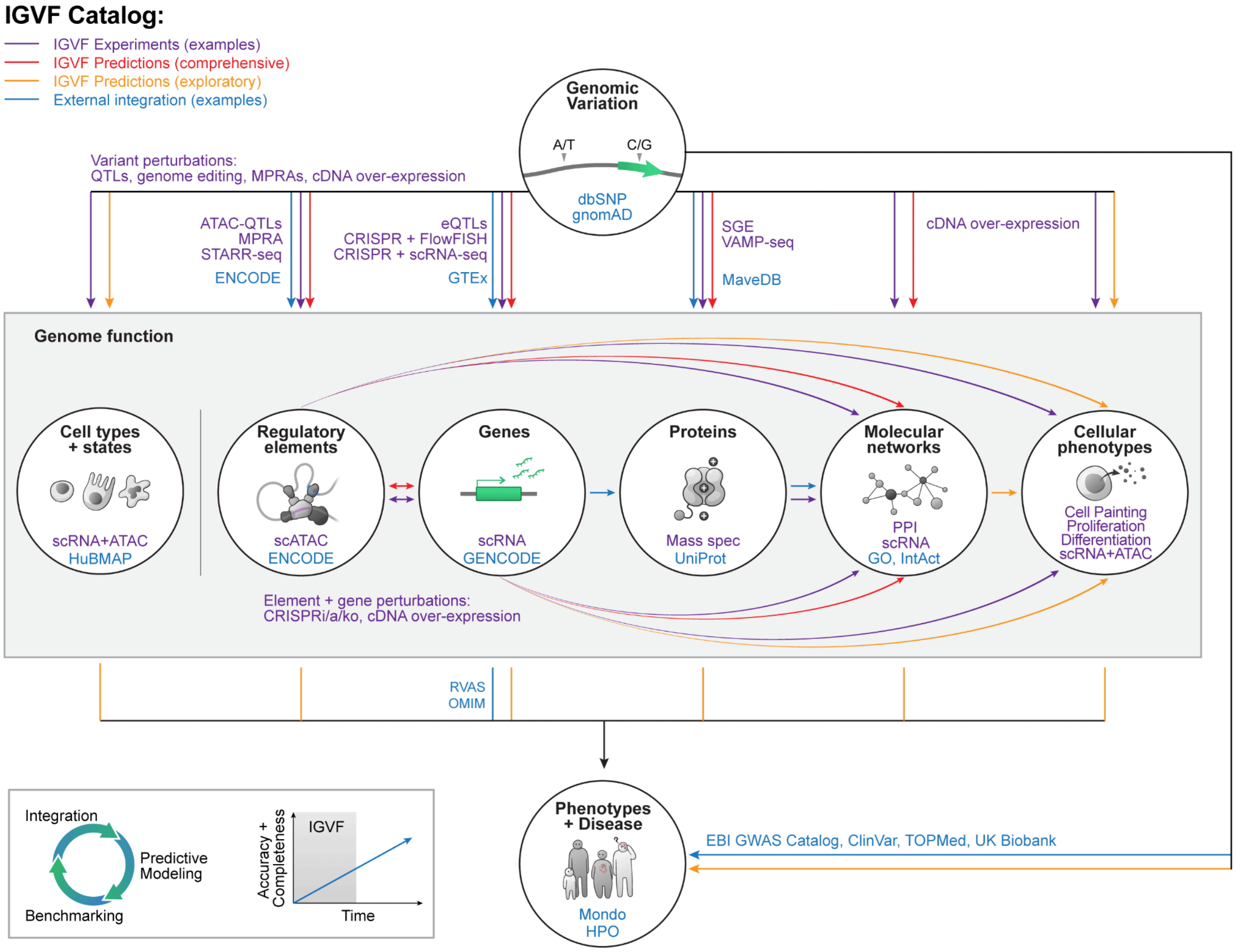

**Figure 3.** The IGVF Catalog of genome function and the impact of genomic variation. IGVF will create a catalog linking genomic variation (top) to genome function (middle box) to phenotype (bottom). Purple: Examples of experimental methods applied by IGVF. Red: Relationships where IGVF plans to develop and apply computational models to comprehensively annotate all possible single-nucleotide variants across many cell types. Orange: Relationships where IGVF plans to develop and apply computational methods in a more targeted fashion, for example in the context of certain cellular phenotypes or diseases. Blue: Examples of external resources or ontologies that could interact with and/or be incorporated into this catalog. Abbreviations and citations: dbSNP[93], gnomAD[4], ENCODE[94], GTEx[17], saturation genome editing (SGE)[22], Variant Abundance by Massively Parallel sequencing (VAMP-seq)[21], MaveDB[23], HuBMAP[19], GENCODE[95], UniProt[96], Gene Ontology (GO)[97], IntAct Molecular Interaction Database[98], Mondo Disease Ontology[99], Human Phenotype Ontology (HPO)[100], rare variant association studies (RVAS), Online Mendelian Inheritance in Man (OMIM)[101].

## Exploring the impact of genomic variation and function on disease

The map-perturb-predict framework and the resulting variant-effect maps will provide new resources for the community to study the impact of genomic variation on human diseases and phenotypes, but this goal presents additional challenges.

For many diseases, an individual's risk is likely to be determined by a combination of thousands of independently acting variants[102,103] — including for many diseases presumed to follow Mendelian inheritance patterns, where penetrance and expressivity may include a component of polygenic risk[104]. Molecular networks are highly interconnected — a single variant may influence multiple genes, multiple gene networks, in diverse cell types — making it difficult to determine which of those genes, networks, and cell types are important for disease[1,17,69,70]. Disease susceptibility can involve many different cell types, possibly at specific timepoints, with effects accumulating over decades or in specific environmental contexts[105]. The impact of genomic variation on genome function can also differ across age, sex, populations, and ancestry: expanding human genetic studies across diverse populations has revealed examples where additional disease associations are discovered due to differences in allele frequencies[106], and some cases in which variants with comparable allele frequencies appear to have different effect sizes on a disease[107–111].

Toward addressing some of these challenges, we will focus on assessing the impacts of variants in molecular networks and diverse cell contexts and then explore how best to apply this framework to: (i) inform clinical variant interpretation, particularly for rare diseases; (ii) learn about molecular and cellular mechanisms underlying risk for common and rare diseases; and (iii) ensure that lessons about the impact of genomic variation on genome function are applicable across diverse populations. Notably, each of these questions represents a major research area involving many strategies beyond those pursued in IGVF[7–9,28,112–114], and these exploratory efforts will seek to integrate with other efforts in the field.

### *Interpreting genomic variation to inform genetic diagnosis*

One key use case for variant-effect maps generated by IGVF, particularly for coding variation, will be to inform the clinical interpretation of single-nucleotide VUS in genes with known and suspected links to Mendelian genetic diseases. Indeed, prior work has shown how applying multiplexed assays of variant effect to study individual genes has been translated into powerful evidence for clinical variant interpretation, for example moving 50% of VUS in *BRCA1*, 70% in *TP53*, 74% in *MSH2*, and 90% in *DDX3X* into more definitive pathogenic or benign classifications[58,115,116]. In some cases, updated genetic test results were provided to individuals that had received VUS results when tested for cancer risk, and diagnostic odysseys were ended for families with DDX3X-associated neurodevelopmental disease.

To expand this approach to additional diseases, IGVF labs will experimentally measure the effects of thousands of variants in known disease genes, with a particular focus on those where identification of loss-of-function variants is clinically actionable[117,118]. We will then assess the extent to which either these experimental data, or computational predictions of variant impact, enrich for variants previously classified as either pathogenic or benign, and determine whether they can be used to calibrate predictors for clinical applications[119]. These variant-effect maps could ultimately substantially reduce the VUS burden in etiological diagnosis of rare disease[114]. Integration of maps for both coding and noncoding variants could also aid in the development of the next-generation polygenic risk score methodologies for better risk characterization in complex phenotypes[107].

*Identifying molecular and cellular mechanisms of disease risk*

Improved variant-effect maps could be transformative for identifying new biological mechanisms that influence genetic risk for disease. In particular, we will seek to understand how best to combine the map-perturb-predict framework and variant-effect maps with human genetic data to nominate variants, genes, cell types, and cellular programs that influence disease risk.

We will study specific diseases and traits, including lipid traits, hematological traits, autoimmune diseases such as systemic lupus erythematosus, cardiometabolic diseases such as coronary artery disease, and neurodegenerative diseases such as Alzheimer's disease. As one example, IGVF investigators are studying variants associated with lipid traits, where GWAS and whole-exome sequencing studies have already identified hundreds of associated noncoding and coding variants, and where certain key genetic pathways involved in lipid handling are already known[11,120–122]. By conducting CRISPR screens to identify variants and regulatory elements that affect lipid uptake in cellular models, testing variant effects on enhancer activity in massively parallel reporter assays, and applying state-of-the-art predictive models, we will evaluate which combinations of experiments and/or predictive models provide the strongest enrichment for disease-associated variation and known causal genes. These combined efforts will help to inform mechanisms of genetic risk for selected diseases, and help to develop strategies to identify causal variants, genes, and pathways for any complex disease.

*Assessing the impact of variation across populations*

IGVF aims to ensure that insights about the impact of genomic variation and genome function are applicable to and inclusive of people of diverse groups. To do so, we will promote diversity in its functional genomics studies, experimentally study and computationally annotate variants observed in diverse populations, study diseases disproportionately affecting disadvantaged or under-represented populations, and explore the extent to which particular variants might exert the same or different effects due to interactions with genetic background or environment[123–125].

We will employ a multi-pronged approach encompassing experimental and computational strategies to achieve its goals. In the current design stage, we have incorporated variants, elements, and genes from diverse populations, including those with differential effects on disease. Biological models will include human iPSCs derived from individuals from different ancestries, and genetically diverse mouse lines from the Collaborative Cross[126]. Finally, certain predictive models, such as those linking noncoding variants to chromatin state and gene expression[62,64,65], can make predictions of variant effects based on measurements obtained in a specific individual, allowing systematic annotation and comparison of variant effects across individuals and groups. Altogether, the data and genome-wide variant-effect maps generated by IGVF will offer insights into variant effects across groups and provide a valuable resource for investigating the effects of variants discovered in diverse populations.

## Data release and resources

A major goal of IGVF is to catalyze future research to understand the relationships between genome function, genomic variation, and phenotype, including for biomedical researchers across diverse disciplines and with diverse needs. To do so, we will build the IGVF Data Resource to enable researchers to easily access and apply IGVF datasets, predictions, and methods (https://igvf.org).

For researchers who want to explore IGVF data and predictions about genome function and variant effects, we will create the IGVF Catalog. The IGVF Catalog will enable searching for information about specific variants, genomic loci, or genes, and will draw from processed data, analysis products, and computational predictions generated by IGVF as well as external data sources such as dbSNP[127], FAVOR[128], gnomAD[4], GENCODE[95], topLD[129], ENCODE[94], GTEx[17], and MaveDB[23] (**Fig. 3**). The IGVF Catalog will be updated several times per year, and all releases will be numbered and archived to maintain reproducibility of studies that depend on its earlier versions. To support users who want to programmatically access IGVF data or analysis products — for example to perform integrative analyses or to develop novel web applications — the IGVF Catalog will also provide a fully featured application programming interface (API) to the underlying knowledge graph.

For researchers who want to access raw or processed data generated by IGVF, we will develop the IGVF Data Portal. The Data Portal will provide web-browser and programmatic access to uniformly processed IGVF datasets, analysis products, and rich metadata, which we anticipate will be useful for users who aim to develop new data analysis methods or predictive models, analyze IGVF data in new ways, or compare their data to IGVF standards. The IGVF Data Portal will follow principles of making data Findable, Accessible, Interoperable, and Reusable (FAIR)[130]. IGVF data and predictions will be made available once they meet pre-defined experimental and computational quality standards. Data will be stored in cloud file buckets to facilitate computing on the data in place and without the need to download data to local servers. Some IGVF data may not have consent for public sharing; such data will be deposited in NHGRI's Analysis, Visualization and Informatics Lab (AnVIL) platform to provide access control in adherence to NIH Policy[131].

For researchers who want to apply IGVF methods and strategies to additional systems, the Data Portal will also share documentation on IGVF standards, protocols, and best practices for experimental design, data analysis, and predictive modeling. These resources will include computational methods, data formats, and consensus data processing pipelines for key assays and analysis products, such as for single-nucleus RNA-seq and ATAC-seq, CRISPR-based experiments, massively parallel reporter assays, eQTL studies, and others. All data processing code will be released with open-source licenses to enable others to analyze similar data in an identical fashion, and we will strive to make sure that it can be run on compute resources accessible to researchers throughout the global research community.

Finally, for all researchers, we will provide training and support on how to access these IGVF resources. To teach researchers how to find and view IGVF data, we will create instructional streaming videos that we will distribute via the IGVF YouTube channel (https://www.youtube.com/@igvf). To teach users how to access data programmatically and use common analytic tools, we will create online notebooks and tutorials demonstrating key uses of the IGVF Portal and the IGVF Catalog. As an additional channel for users to interact with the IGVF Consortium, we will host interactive online seminars and webinars.

Altogether, we expect that these resources will enable a wide range of scientific activities, expanding far beyond the specific studies undertaken by the IGVF Consortium.

## Collaborations and community

Toward advancing our collective efforts to understand genomic variation and genome function — a grand challenge that demands global and interdisciplinary collaboration — IGVF welcomes collaboration with and input from the broader scientific community. Researchers interested in joining IGVF can apply for Affiliate Membership. Affiliate Membership allows investigators to fully participate in working groups and other IGVF collaborations, and thereby help drive the vision, goals, and execution of consortium activities. For more information, visit https://igvf.org/affiliate-membership/.

IGVF is actively coordinating with other consortia, including ClinGen[8], the Genomics Research to Elucidate the Genetics of Rare diseases (GREGoR) consortium, and the Atlas of Variant Effects (AVE) Alliance[132]. These collaborations will facilitate the open exchange and interoperability of genomic data and resources, for example to use common variant naming schema, genome and transcriptome builds, and experimental and analysis pipelines.

Similarly, IGVF data and analysis products will benefit from close interactions with efforts to characterize human genomic variation and assemblies, such as the Human Pangenome Reference Consortium (HPRC)[133]; with efforts to catalog disease-associated variation across ancestries, including All of Us[134], TOPMed[10], and other biobanks; with efforts to map the activities of variants, regulatory elements, and genes at single-cell resolution, such as the Human Cell Atlas[20] and HuBMAP[19]; and with efforts to compare and evaluate strategies for interpreting genetic variation associated with disease, such as the International Common Disease Alliance[28]. Strong collaborative ties among such efforts and IGVF will propel scientific advances that shape how both basic and clinical research are performed.

## Outlook and Perspectives

With the rapid expansion of human genetics studies linking variation to disease, the interpretation of the impact of genomic variation on function is currently a rate-limiting step for delivering on the promise of precision medicine. The IGVF Consortium will deploy a coordinated strategy for accelerating progress, including generating large-scale data resources, predictive models, and initial variant-effect maps that will reveal new insights into how genomic variation impacts function and phenotype. The tools, data resource, and strategy developed by IGVF — including new experimental assays, design strategies, predictive models, computational methods, data sharing standards, and more — will provide a foundation to facilitate future efforts. We will prioritize open data and resource sharing, inclusion, and outreach so that all members of the research community can participate and benefit.

While ambitious, IGVF activities do have limitations in scope, and many challenges lie ahead. Genomic technologies, both experimental and computational, are developing rapidly, and balancing the implementation of the newest scalable tools with continuing standards to ensure data interoperability will require attention. While data generation technologies have increased throughput exponentially over the last 15 years, the amount of data needed to build accurate models of genome function is unknown, and fully realizing the goal of mapping the impact of genomic variation on function will require additional advances in both experimental and computational methods. In particular, the development of computational methods to predict synergistic interactions among variants, environments, and time spans of effects that can occur over decades are open problems. We will initially focus on specific biological systems and cellular models according to its members' expertise, but full exploration of the many cell types and disease areas relevant to human biology will require community efforts. IGVF aims for systematic analysis of certain aspects of genome function — gene regulation, protein function, and molecular networks. Additional work is required to explore other important layers of genome function, including effects on nuclear organization and chromatin compartmentalization; RNA splicing, transport, and translation; and impacts on cellular phenotypes, cell-cell interactions, and communication. For all of these challenges, the framework developed by the IGVF Consortium to develop and benchmark methods, refine best practices and standards, and share data and methods will drive scientific discoveries in human health and disease for years to come.

## Author List

***Writing group* (ordered by contribution):** Jesse M. Engreitz, Heather A. Lawson, Harinder Singh, Lea M. Starita, Gary C. Hon, Hannah Carter, Nidhi Sahni, Timothy E. Reddy, Xihong Lin, Yun Li, Nikhil V. Munshi, Maria H. Chahrour, Alan P. Boyle, Benjamin C. Hitz, Ali Mortazavi, Mark Craven, Karen L. Mohlke, Luca Pinello, Ting Wang

***Steering Committee Co-Chairs* (alphabetical by last name):** Anshul Kundaje, Karen L. Mohlke, Feng Yue

***Working Group and Focus Group Co-Chairs* (alphabetical by last name):**
**Catalog:** Michael I. Love, Lea M. Starita, Feng Yue
**Characterization:** Gary C. Hon, Martin Kircher, Timothy E. Reddy
**Computational Analysis, Modeling, and Prediction:** Xihong Lin, Jian Ma, Predrag Radivojac
**Project Design:** Brunilda Balliu, Jesse M. Engreitz, Nidhi Sahni
**Mapping:** Nina P. Farrell, Brian A. Williams
**Networks:** Hannah Carter, Danwei Huangfu
**Standards and Pipelines:** Anshul Kundaje, Luca Pinello
**Cardiometabolic:** Nikhil V. Munshi, Chong Y. Park, Thomas Quertermous
**Cellular Programs and Networks:** Hannah Carter, Jishnu Das
**Coding Variants:** Michael A. Calderwood, Douglas M. Fowler, Predrag Radivojac, Lea M. Starita, Marc Vidal
**CRISPR:** Lucas Ferreira, Luca Pinello
**Defining and Systematizing Function:** Mark Craven, Sean D. Mooney, Vikas Pejaver
**Enumerating Variants:** Benjamin C. Hitz, Jingjing Zhao
**Evolution:** Steven Gazal, Evan Koch, Steven K. Reilly, Shamil Sunyaev
**Imaging:** Anne E. Carpenter,
**Immune:** Jason D. Buenrostro, Christina S. Leslie, Rachel E. Savage
**Impact on Diverse Populations:** Stefanija Giric, Yun Li
**iPSC:** Chongyuan Luo, Kathrin Plath
**MPRA:** Alejandro Barrera, Michael I. Love, Max Schubach
**Noncoding Variants:** Jesse M. Engreitz, Jill E. Moore, Nidhi Sahni
**Neuro:** Nadav Ahituv, Maria H. Chahrour
**Phenotypic Impact and Function:** Kushal Dey, Xihong Lin
**QTL/Statgen:** Brunilda Balliu, Ingileif Hallgrimsdottir, Kyle Gaulton, Saori Sakaue
**Single Cell:** Sina Booeshaghi, Anshul Kundaje, Eugenio Mattei, Ali Mortazavi, Surag Nair, Lior Pachter, Austin Wang

**Characterization Awards (contact PI, MPIs (alphabetical by last name), other members (alphabetical by last name)):**
***UM1HG011966***: Jay Shendure[1,5,171,172,173], Nadav Ahituv[2], Martin Kircher[3,4], Vikram Agarwal[1,174], Andrew Blair[2], Theofilos Chalkiadakis[4], Florence M. Chardon[1], Pyaree M Dash[4], Chengyu Deng[2], Nobuhiko Hamazaki[1], Pia Keukeleire[3], Connor Kubo[1], Jean-Benoît Lalanne[1], Thorben Maass[3], Beth Martin[1], Troy McDiarmid[1], Mai Nobuhara[2], Nicholas F Page[2], Sam Regalado[1], Max Schubach[4], Jasmine Sims[2], Aki Ushiki[2], Jingjing Zhao[2]

***UM1HG011969***: Lea M. Starita[1,5], Douglas M. Fowler[1,5], Sabrina M. Best[1], Gabe Boyle[1], Nathan Camp[6], Silvia Casadei[1], Estelle Y. Da[7], Moez Dawood[5,8], Samantha C. Dawson[6], Shawn Fayer[1], Audrey Hamm[1], Richard G. James[6], Gail P. Jarvik[1], Abbye E. McEwen[1,5,9], Nick Moore[7], Lara A. Muffley[1], Sriram Pendyala[1], Nicholas A. Popp[1], Mason Post[1], Alan F. Rubin[7], Jay Shendure[1,5,171,172,173], Nahum T. Smith[1], Alan F. Rubin[1], Jeremy Stone[5], Malvika Tejura[1], Ziyu R. Wang[1], Melinda K. Wheelock[1], Ivan Woo[1], Brendan D. Zapp[1]

***UM1HG011972:*** Jesse M. Engreitz[10,11,12,61], Thomas Quertermous[13], Dulguun Amgalan[10,11], Aradhana Aradhana[10], Sophia M. Arana[10], Michael C. Bassik[10], Julia R. Bauman[10], Asmita Bhattacharya[10],

Xiangmeng Shawn Cai[10,11,22], Ziwei Chen[14], Stephanie Conley[10,11], Salil Deshpande[15], Benjamin R. Doughty[10], Peter P. Du[10], Casey Gifford[10,11,16,17], William J. Greenleaf[10,161], Andreas R. Gschwind[10], Katherine Guo[10,11], Sarasa Isobe[18], Evelyn Jagoda[12,13], Nimit Jain[10], Hank Jones[10,11], Helen Y. Kang[10,11], Samuel H. Kim[162], YeEun Kim[163], Sandy Klemm[10], Anshul Kundaje[10,14], Soumya Kundu[14], Mauro Lago-Docampo[18], Yannick C. Lee-Yow[10,11], Roni Levin-Konigsberg[10], Daniel Y. Li[13], Dominik Lindenhofer[19], X. Rosa Ma[10,11], Georgi K. Marinov[10], Gabriella E. Martyn[10,11], Eyal Metzl-Raz[10], Joao P. Monteiro[13], Michael T. Montgomery[10,11], Kristy S. Mualim[11,20,164], Chad Munger[10,11], Glen Munson[12], Tri C. Nguyen[10,11], Trieu Nguyen[13], Brian T. Palmisano[13], Anusri Pampari[14], Chong Y. Park[13], Marlene Rabinovitch[18], Markus Ramste[13], Judhajeet Ray[12], Kevin R. Roy[10,21], Oriane M. Rubio[11], Julia M. Schaepe[22], Gavin Schnitzler[13], Jacob Schreiber[10], Disha Sharma[13], Maya U. Sheth[10,11,22], Huitong Shi[13], Vasundhara Singh[12], Riya Sinha[23], Lars M. Steinmetz[10,19,21], Jason Tan[10,11,14], Anthony Tan[10,11], Josh Tycko[10], Raeline C. Valbuena[10], Valeh Valiollah Pour Amiri[10], Mariëlle J.F.M. van Kooten[10], Alun Vaughan-Jackson[10], Anthony Venida[10], Chad S. Weldy[13], David Yao[10], Tony Zeng[10,11], Ronghao Zhou[10,11]

***UM1HG011989:*** Marc Vidal[24,25], Michael A. Calderwood[24,25,26], Anne E. Carpenter[27], Beth A. Cimini[27], Georges Coppin[24,25,26,28], Atina G. Coté[29,30,31], Marzieh Haghighi[27], Tong Hao[24,25,26], David E. Hill[24,25,26], Jessica Lacoste[29,30], Florent Laval[24,25,26,28,184], Chloe Reno[29,30], Frederick P. Roth[29,30,31,32], Shantanu Singh[27], Kerstin Spirohn-Fitzgerald[24,25], Mikko Taipale[29,30], Tanisha Teelucksingh[29], Maxime Tixhon[24,25,26,185], Anupama Yadav[24,25,26], Zhipeng Yang[24,25,26]

***UM1HG011996:*** Gary C. Hon[33,34,35], W. Lee Kraus[33,34], Nikhil V. Munshi[36,37], Daniel A. Armendariz[33], Maria H. Chahrour[38,211,212,213,214], Ashley E. Dederich[39], Lauretta El Hayek[38], Sean C. Goetsch[36], Kiran Kaur[38], Hyung Bum Kim[33], Melissa K. McCoy[39], Mpathi Z. Nzima[33], Carlos A. Pinzón-Arteaga[40], Bruce A. Posner[39], Daniel A. Schmitz[37], Sushama Sivakumar[36,37], Anjana Sundarrajan[33], Lei Wang[33], Yihan Wang[33], Jun Wu[37], Lin Xu[40,41], Jian Xu[42], Leqian Yu[37], Yanfeng Zhang[40], Huan Zhai[33], Qinbo Zhou[40]

***UM1HG012003:*** Hyejung Won[43,44], Michael I. Love[43,204], Karen L. Mohlke[43], Jessica L. Bell[43,44], K. Alaine Broadaway[43], Katherine N. Degner[43,44], Amy S. Etheridge[43], Stefanija Giric[43], Beverly H. Koller[43], Yun Li[43,204], Won Mah[43,44], Wancen Mu[204], Kimberly D. Ritola[44,205], Jonathan D. Rosen[43], Sarah A. Schoenrock[43,44], Rachel A. Sharp[43,44]

***UM1HG012010:*** Luca Pinello[45,61], Daniel Bauer[47,48], Guillaume Lettre[49,50], Richard Sherwood[51], Basheer Becerra[47,48], Logan J. Blaine[45,52], Lucas Ferreira[52,53], Matthew J. Francoeur[51], Ellie N. Gibbs[51], Nahye Kim[45,54,217], Emily M. King[45,54,217,218], Benjamin P. Kleinstiver[45,54,217], Estelle Lecluze[49], Zhijian Li[45,46], Zain M. Patel[45,46], Quang Vinh Phan[51], Jayoung Ryu[45,52], Marlena L Starr[53], Ting Wu[48,53]

***UM1HG012053:*** Charles A. Gersbach[55,56], Gregory E. Crawford[56,57], Timothy E. Reddy[58], Andrew S. Allen[58], William H. Majoros[58], Nahid Iglesias[55,56], Alejandro Barrera[56,58], Ruhi Rai[56], Revathy Venukuttan[56], Boxun Li[55,56], Taylor Anglen[56,59], Lexi R. Bounds[55,56], Marisa C. Hamilton[56], Siyan Liu[56], Sean R. McCutcheon[55,56], Christian D. McRoberts Amador[56,60], Samuel J. Reisman[56,59], Maria A. ter Weele[55,56], Josephine C. Bodle[55,56], Helen L. Streff[55,56], Keith Siklenka[58], Kari Strouse[58]

**Mapping Awards (contact PI, MPIs (alphabetical by last name), other members (alphabetical by last name)):**
***UM1HG011986:*** Jason D. Buenrostro[61,62], Bradley E. Bernstein[61,63], Juliana Babu[61,62], Guillermo Barreto Corona[61], Kevin Dong[61], Fabiana M. Duarte[61,62], Neva C. Durand[61], Charles B. Epstein[61], Kaili Fan[61,62,98], Nina P. Farrell[61], Elizabeth Gaskell[61], Amelia W. Hall[61], Alexandra M. Ham[61], Mei K. Knudson[61], Eugenio Mattei[61], Rachel E. Savage[61,62], Noam Shoresh[61], Siddarth Wekhande[61], Cassandra M. White[61], Wang Xi[61,62]

***UM1HG012076:*** Ansuman T. Satpathy[64,65,66], M. Ryan Corces[73,74,187], Serena H. Chang[73,74,187], Iris M. Chin[73,74,187], James M. Gardner[75,76], Zachary A. Gardell[73,74,187], Jacob C. Gutierrez[64,66], Alia W. Johnson[73,74,187], Lucas Kampman[73,74,187], Maya Kasowski[64,72], Caleb A. Lareau[64,65,66], Vincent Liu[64,66], Leif S. Ludwig[67,68], Christopher S. McGinnis[64,65,66], Shreya Menon[73,74,187], Adam W. Turner[73,74,187], Chun J. Ye[69,70,71], Yajie Yin[64,66], Wenxi Zhang[64]

*UM1HG012077:* Ali Mortazavi[188,189], Barbara J. Wold[190,191], Sina Booeshaghi[190], Maria Carilli[192], Dayeon Cheong[188], Ghassan Filibam[188], Kim Green[188,193], Ingileif Hallgrimsdottir[190], Shimako Kawauchi[189], Charlene Kim[190], Heidi Liang[189], Rebekah Loving[190], Laura Luebbert[190], Grant MacGregor[188], Angel G Merchan[190], Lior Pachter[190,194], Elisabeth Rebboah[188], Fairlie Reese[188,189], Narges Rezaie[188,189], Jasmine Sakr[189,195], Delaney Sullivan[190], Nikki Swarna[192], Diane Trout[190], Sean Upchurch[190], Ryan Weber[188], Brian A. Williams[190]

**Predictive Modeling Awards (contact PI, MPIs (alphabetical by last name), other members (alphabetical by last name)):**
*U01HG011952:* Alan P. Boyle[196,197], Christopher P. Castro[196], Elysia Chou[196], Fan Feng[196], Andre Guerra[198], Yuanhao Huang[196], Linghua Jiang[196], Jie Liu[196], Ryan E. Mills[196,197], Weizhou Qian[196], Tingting Qin[196], Maureen A. Sartor[196,198], Rintsen N. Sherpa[196], Jinhao Wang[196], Yiqun Wang[196], Joshua D. Welch[196], Zhenhao Zhang[196], Nanxiang Zhao[196]

*U01HG011967:* Andrew S. Allen[58], Sayan Mukherjee[77,78,79], C. David Page[58], Shannon Clarke[58], Richard W. Doty[58], Yuncheng Duan[80], Raluca Gordan[58,79], Kuei-Yueh Ko[58], Shengyu Li[58], Boyao Li[58], William H. Majoros[58], Timothy E. Reddy[58], Alexander Thomson[58]

*U01HG012009:* Soumya Raychaudhuri[51,82], Alkes Price[83,84,82], Shamil Sunyaev[51,52], Thahmina A. Ali[81], Kushal K. Dey[81,83], Arun Durvasula[83,85], Manolis Kellis[46,220], Evan Koch[52], Saori Sakaue[51,82]

*U01HG012022:* Predrag Radivojac[86], Lilia M. Iakoucheva[87], Tulika Kakati[87], Sean D. Mooney[88], Yile Chen[88], Mariam Benazouz[88], Vikas Pejaver[89,90], Shantanu Jain[86,215], Daniel Zeiberg[86], M. Clara De Paolis Kaluza[86], Michelle Velyunskiy[86]

*U01HG012039:* Mark Craven[91], Audrey Gasch[92], Kunling Huang[93], Yiyang Jin[91], Qiongshi Lu[91], Jiacheng Miao[91], Michael Ohtake[94], Eduardo Scopel[92], Robert D. Steiner[95,96,97], Yuriy Sverchkov[91]

*U01HG012064:* Zhiping Weng[98], Manuel Garber[98], Xihong Lin[84,100], Yu Fu[98], Natalie Haas[98], Xihao Li[43,84,204], Nishigandha Phalke[98], Shuo C. Shan[98], Nicole Shedd[98], Eric Van Buren[84], Tianxiong Yu[98], Yi Zhang[101], Hufeng Zhou[84]

*U01HG012064:* Anshul Kundaje[10,14], Alexis Battle[102,103,104,105], Ziwei Chen[14], Salil Deshpande[15], Jesse M. Engreitz[10,11,12,61], Livnat Jerby[10], Eran Kotler[10], Soumya Kundu[10,14], Andrew R. Marderstein[64], Georgi K. Marinov[10], Stephen B. Montgomery[10,64,106], Surag Nair[14], AkshatKumar Nigam[10,14], Evin M. Padhi[64], Anusri Pampari[14], Aman Patel[14], Jonathan Pritchard[10], Ivy Raine[10], Vivekanandan Ramalingam[10], Kameron Rodrigues[64], Jacob M. Schreiber[10], Arpita Singhal[14], Riya Sinha[15], Valeh Valiollah Pour Amiri[10], Austin T. Wang[14]

**Network Projects (contact PI, MPIs (alphabetical by last name), other members (alphabetical by last name)):**
*U01HG012041:* Harinder Singh[107], Jishnu Das[107], Nidhi Sahni[108,109,110], Marisa Abundis[111], Deepa Bisht[112], Trirupa Chakraborty[107], Jingyu Fan[107], David R. Hall[107], Zarifeh H. Rarani[107], Abhinav Jain[112], Babita Kaundal[112], Swapnil Keshari[107], Daniel McGrail[113,114], Nicholas A. Pease[107], Vivian F. Yi[107], S. Stephen Yi[115,116]

*U01HG012047:* Hao Wu[117], Sreeram Kannan[118], Hongjun Song[119], Jingli Cai[120], Ziyue Gao[117], Ronni Kurzion[119], Julia I. Leu[117], Fan Li[117], Dongming Liang[117], Guo-li Ming[119], Kiran Musunuru[120], Qi Qiu[117], Junwei Shi[121], Yijing Su[119], Sarah Tishkoff[117], Ning Xie[117], Qian Yang[119], Wenli Yang[120], Hongjie Zhang[117], Zhijian Zhang[119]

*U01HG012051:* Danwei Huangfu[122,123], Michael A. Beer[124], Ronald Cutler[125], Rachel A. Glenn[122,123,126], Renhe Luo[122,123], Jin Woo Oh[124], Milad Razavi-Mohseni[124], Dustin Shigaki[124], Simone Sidoli[125], Thomas Vierbuchen[122,123], Jielin Yan[122,123], Yunxiao Yang[124]

***U01HG012059:*** Maike Sander[127], Hannah Carter[128], Kyle J. Gaulton[127], Bing Ren[129,130], Weronika Bartosik[129], Hannah S. Indralingam[129], Adam Klie[131], Hannah Mummey[131], Mei-Lin Okino[132], Gaowei Wang[127], Nathan R. Zemke[129],Kai Zhang[129], Han Zhu[127]

***U01HG012079:*** Chongyuan Luo[133], Kathrin Plath[134], Noah Zaitlen[135], Brunilda Balliu[136,137,138], Jason Ernst[134,137], Justin Langerman[134], Terence Li[133], Yu Sun[134]

***U01HG012103:*** Christina S. Leslie[199], Alexander Y. Rudensky[200,201], Preethi K. Periyakoil[199], Vianne R. Gao[199], Melanie H. Smith[202], Norman M. Thomas[199], Laura T. Donlin[202,203], Amit Lakhanpal[202], Kaden M. Southard[199], Rico C. Ardy[199]

**Data and Administrative Coordinating Center Awards (contact PI, MPIs (alphabetical by last name), other members (alphabetical by last name)):**
***U24HG012103:*** J. Michael Cherry[10], Mark B. Gerstein[166,167,168,169,170], Kalina Andreeva[10], Pedro R. Assis[10], Beatrice Borsari[166,167], Eric Douglass[10], Shengcheng Dong[10], Idan Gabdank[10], Keenan Graham[10], Benjamin C. Hitz[10], Otto Jolanki[10], Jennifer Jou[10], Meenakshi S. Kagda[10], Jin-Wook Lee[10], Mingjie Li[10], Khine Lin[10], Stuart R. Miyasato[10], Joel Rozowsky[166,167], Corinn Small[10], Emma Spragins[10], Forrest Y. Tanaka[10], Ian M. Whaling[10], Ingrid A. Youngworth[10], Cricket A. Sloan[10]

***U24HG012103:*** Ting Wang[139,140], Feng Yue[175,176], Eddie Belter[140], Xintong Chen[175], Rex L. Chisholm[178], Sarah Cody[140], Patricia Dickson[180], Changxu Fan[139], Lucinda Fulton[140], Heather A. Lawson[139], Daofeng Li[139], Tina Lindsay[140], Yu Luan[175], Yuan Luo[179], Huijue Lyu[175], Xiaowen Ma[139], Jian Ma[165], Juan Macias-Velasco[139], Karen H. Miga[186], Kara Quaid[139], Nathan Stitziel[181], Barbara E. Stranger[177], Chad Tomlinson[140], Juan Wang[175], Wenjin Zhang[139], Bo Zhang[182], Guoyan Zhao[139,183,216], Xiaoyu Zhuo[139]

**IGVF Affiliate Member Projects (Contact PIs, other members (alphabetical by last name)):**
Kristen Brennand[219]
Alberto Ciccia[210], Samuel B. Hayward[210], Jen-Wei Huang[210], Giuseppe Leuzzi[210], Angelo Taglialatela[210], Tanay Thakar[210], Alina Vaitsiankova[210]
Kushal K. Dey[46,141], Thahmina A. Ali[141]
Steven Gazal[142,143,144], Artem Kim[142]
H. Leighton Grimes[209], Nathan Salomonis[209]
Rajat Gupta[13], Shi Fang[13], Vivian Lee-Kim[13]
Matthias Heinig[145,146,147], Corinna Losert[145,146]
Thouis R. Jones[12], Elisa Donnard[12], Maddie Murphy[12], Elizabeth Roberts[12], Susie Song[12]
Jill E. Moore[98]
Sara Mostafavi[221,222], Alexander Sasse[221], Anna Spiro[221]
Len A. Pennacchio[148,151], Momoe Kato[148], Michael Kosicki[148], Brandon Mannion[148], Neil Slaven[148]
Axel Visel[148,151]
Katherine S. Pollard[152,153,154], Siron Drusinsky[152,153], Sean Whalen[152]
John Ray[1,172,208], Ingrid A. Harten[172], Ching-Huang Ho[172]
Steven K. Reilly[223]
Neville E. Sanjana[149,150], Christina Caragine[149,150], John A. Morris[149,150]
Davide Seruggia[155,156], Ana Patricia Kutschat[155,156], Sandra Wittibschlager[155,156]
Han Xu[108], Rongjie Fu[108], Wei He[108], Liang Zhang[108]
S. Stephen Yi[157,158], Daniel Osorio[157,158]

**NHGRI Program Management (alphabetical by last name):** Zo Bly[159], Stephanie Callouri [160,206], Daniel A. Gilchrist[160], Carolyn M. Hutter[160], Stephanie A. Morris[160], Michael J. Pazin[160], Ella K. Samer[160,207]

## Affiliations:

1. Department of Genome Sciences, University of Washington, Seattle, WA, USA
2. Department of Bioengineering and Therapeutic Sciences, Institute for Human Genetics, University of California San Francisco, San Francisco, California, USA

3. Institute of Human Genetics, University Medical Center Schleswig-Holstein, University of Lübeck, 23562 Lübeck, Germany
4. Exploratory Diagnostic Sciences, Berlin Institute of Health at Charité-Universitätsmedizin Berlin, 10117 Berlin, Germany
5. Brotman Baty Institute for Precision Medicine, Seattle, WA., USA
6. Center of immunotherapy and Immunity, Seattle Children's Research Institute, Seattle, WA, USA
7. Bioinformatics Division, WEHI, Parkville, VIC, Australia
8. Human Genome Sequencing Center, Baylor College of Medicine, Houston, TX, USA
9. Department of Laboratory Medicine and Pathology, University of Washington, Seattle, WA, USA
10. Department of Genetics, Stanford University School of Medicine, Stanford, CA, USA
11. Basic Science and Engineering Initiative, Stanford Children's Health, Betty Irene Moore Children's Heart Center, Stanford, CA, USA
12. The Novo Nordisk Foundation Center for Genomic Mechanisms of Disease, Broad Institute of MIT and Harvard, Cambridge, MA, USA
13. Division of Cardiovascular Medicine, School of Medicine, Stanford University
14. Department of Computer Science, Stanford University School of Medicine, Stanford, CA, USA
15. Institute for Computational and Mathematical Engineering, Stanford University, Stanford, CA, USA
16. Department of Pediatrics, Stanford University School of Medicine, Stanford, CA, USA
17. Institute for Stem Cell Biology and Regenerative Medicine, Stanford University School of Medicine, Stanford, CA, USA
18. Division of Pediatric Cardiology and Cardiovascular Institute, Stanford University School of Medicine, Stanford University
19. European Molecular Biology Laboratory (EMBL), Genome Biology Unit, Heidelberg, Germany
20. Department of Biology, Stanford University, Stanford, CA, USA
21. Stanford Genome Technology Center, Palo Alto, CA., USA
22. Department of Bioengineering, Stanford University School of Engineering, Stanford, CA, USA
23. Department of Biomedical Informatics, Stanford University School of Medicine, Stanford, CA, USA
24. Center for Cancer Systems Biology (CCSB), Dana-Farber Cancer Institute, Boston, MA, USA
25. Department of Genetics, Blavatnik Institute, Harvard Medical School, Boston, MA, USA
26. Department of Cancer Biology, Dana-Farber Cancer Institute, Boston, MA, USA
27. Imaging Platform, Broad Institute of Harvard and MIT, Cambridge, Massachusetts
28. Laboratory of Viral Interactomes, GIGA Institute, University of Liège, Liège, Belgium
29. Donnelly Centre for Cellular and Biomolecular Research (CCBR), University of Toronto, Toronto, Ontario, Canada
30. Department of Molecular Genetics, University of Toronto, Toronto, Ontario, Canada.
31. Lunenfeld-Tanenbaum Research Institute (LTRI), Sinai Health System, Toronto, Ontario, Canada
32. Department of Computer Science, University of Toronto, Toronto, Ontario, Canada
33. Cecil H. and Ida Green Center for Reproductive Biology Sciences, University of Texas Southwestern Medical Center, Dallas, TX, USA
34. Department of Obstetrics and Gynecology, University of Texas Southwestern Medical Center, Dallas, TX, USA
35. Department of Bioinformatics, University of Texas Southwestern Medical Center, Dallas, TX, US
36. Department of Internal Medicine, Division of Cardiology, University of Texas Southwestern Medical Center, Dallas, TX, USA
37. Department of Molecular Biology, University of Texas Southwestern Medical Center, Dallas, TX, USA
38. Eugene McDermott Center for Human Growth and Development, University of Texas Southwestern Medical Center, Dallas, TX, USA
39. Department of Biochemistry, University of Texas Southwestern Medical Center, TX, USA
40. Quantitative Biomedical Research Center, Peter O'Donnell Jr. School of Public Health, University of Texas Southwestern Medical Center, Dallas, TX, U.S.A
41. Department of Pediatrics, Division of Hematology/Oncology, University of Texas Southwestern Medical Center, Dallas, TX, U.S.A

42. Children’s Medical Center Research Institute, University of Texas Southwestern Medical Center, TX, USA
43. Department of Genetics, University of North Carolina at Chapel Hill, Chapel Hill, NC, USA
44. Neuroscience Center, University of North Carolina at Chapel Hill, Chapel Hill, NC, USA
45. Department of Pathology, Harvard Medical School, Boston, MA, USA
46. Broad Institute of MIT and Harvard, Boston, MA, USA
47. Division of Hematology/Oncology, Boston Children's Hospital, Boston, MA, USA
48. Department of Pediatrics, Harvard Medical School, Boston, MA, USA
49. Montreal Heart Institute, Montreal, Quebec, H1T 1C8, Canada
50. Département de Médecine, Université de Montréal, Montréal, Quebec, H3T 1J4, Canada
51. Department of Medicine, Brigham and Women's Hospital and Harvard Medical School, Boston, MA 02115
52. Department of Biomedical Informatics, Harvard Medical School, Boston, MA, USA
53. Division of Hematology/Oncology, Boston Children's Hospital, Boston, MA, USA
54. Center for Genomic Medicine and Department of Pathology, Massachusetts General Hospital, Boston, MA, USA
55. Department of Biomedical Engineering, Duke University, Durham, NC, USA
56. Center for Advanced Genomic Technologies, Duke University, Durham, NC
57. Department of Pediatrics, Duke University, Durham, NC, USA
58. Department of Biostatistics and Bioinformatics, Duke University Medical Center, Durham, NC
59. Department of Cell Biology, Duke University Medical Center, Durham, NC, USA
60. Department of Pharmacology and Cancer Biology, Duke University Medical Center, Durham, NC, USA
61. Gene Regulation Observatory, The Broad Institute of MIT and Harvard, Cambridge, MA, USA
62. Department of Stem Cell and Regenerative Biology, Harvard University, Cambridge, MA, USA
63. Department of Cancer Biology, Dana-Farber Cancer Institute, Boston, MA, USA
64. Department of Pathology, Stanford University, Stanford, CA, USA
65. Parker Institute for Cancer Immunotherapy, San Francisco, CA, United States
66. Gladstone-UCSF Institute of Genomic Immunology, San Francisco, CA, 94158, USA.
67. Berlin Institute of Health at Charité – Universitätsmedizin Berlin, Berlin, Germany
68. Max-Delbrück-Center for Molecular Medicine in the Helmholtz Association (MDC), Berlin Institute for Medical Systems Biology (BIMSB), Berlin, Germany
69. Institute for Human Genetics, Department of Medicine, Division of Rheumatology, University of California, San Francisco, CA, United States
70. Parker Institute for Cancer Immunotherapy, San Francisco, CA, United States
71. Chan Zuckerberg Biohub, San Francisco, CA, United States
72. Sean N Parker Center for Allergy and Asthma Research, Stanford University, Stanford, CA, USA
73. Gladstone Institute of Neurological Disease, San Francisco, CA, USA
74. Department of Neurology, University of California San Francisco, San Francisco, CA, USA
75. Department of Surgery, University of California San Francisco, San Francisco, CA, USA
76. Diabetes Center, University of California San Francisco, San Francisco, CA, USA
77. Department of Statistical Science, Duke University, Durham, NC, USA
78. Department of Mathematics, Duke University, Durham, NC, USA
79. Department of Computer Science, Duke University. Durham, NC, USA
80. Department of Biology, Duke University, Durham NC, USA
81. Computational and Systems Biology Program, Memorial Sloan Kettering Cancer Center, New York NY, USA
82. Department of Medical and Population Genetics, Broad Institute, Cambridge, MA, USA
83. Department of Epidemiology, Harvard T.H.Chan School of Public Health, Boston, MA, USA
84. Department of Biostatistics, Harvard T.H. Chan School of Public Health, Boston, MA, USA
85. Department of Genetics, Harvard Medical School, Boston, MA, USA
86. Khoury College of Computer Sciences, Northeastern University, Boston, MA 02115, USA
87. Department of Psychiatry, University of California San Diego, La Jolla, CA 92093, USA
88. Department of Biomedical Informatics and Medical Education, University of Washington, Seattle, WA 98195, USA
89. Institute for Genomic Health, Icahn School of Medicine at Mount Sinai, New York, NY 10029, USA

90. Department of Genetics and Genomic Sciences, Icahn School of Medicine at Mount Sinai, New York, NY 10029, USA
91. Department of Biostatistics and Medical Informatics, University of Wisconsin, Madison, WI, USA
92. Department of Genetics, University of Wisconsin, Madison, WI, USA
93. Department of Statistics, University of Wisconsin, Madison, WI, USA
94. Department of Computer Sciences, University of Wisconsin, Madison, WI, USA
95. Department of Pediatrics, University of Wisconsin, Madison, WI, USA
96. Prevention Genetics Inc., Marshfield, WI, USA
97. Marshfield Clinic, Marshfield, WI, USA
98. Program in Bioinformatics and Integrative Biology, UMass Chan Medical School, Worcester, MA, USA
99. Department of Data Science, Dana-Farber Cancer Institute, Boston, MA
100. Department of Statistics, Harvard University, Cambridge, MA
101. Department of Data Science, Dana-Farber Cancer Institute, Boston, MA
102. Department of Biomedical Engineering, Johns Hopkins University, Baltimore, MD, USA
103. Malone Center for Engineering in Healthcare, Johns Hopkins University, Baltimore, MD, USA
104. Department of Computer Science, Johns Hopkins University, Baltimore, MD, USA
105. Department of Genetic Medicine, Johns Hopkins University, Baltimore, MD, USA
106. Department of Biomedical Data Science, Stanford University, Stanford, CA, USA
107. Departments of Immunology and Computational and Systems Biology, University of Pittsburgh, Pittsburgh, PA
108. Department of Epigenetics and Molecular Carcinogenesis, University of Texas MD Anderson Cancer Center, Houston, TX, USA
109. Department of Epigenetics and Molecular Carcinogenesis, University of Texas MD Anderson Cancer Center, Houston, TX, USA
110. Department of Bioinformatics and Computational Biology, The University of Texas MD Anderson Cancer Center, Houston, TX, USA
111. Departments of Immunology and Computational and Systems Biology, University of Pittsburgh, Pittsburgh, PA, USA
112. Department of Epigenetics and Molecular Carcinogenesis, University of Texas MD Anderson Cancer Center, Houston, TX, USA
113. Center for Immunotherapy and Precision Immuno-Oncology, Cleveland Clinic, Cleveland, OH, USA
114. Center for Immunotherapy and Precision Immuno-Oncology, Cleveland Clinic, Cleveland, OH, USA
115. Livestrong Cancer Institutes, Department of Oncology, and Department of Biomedical Engineering, The University of Texas at Austin, Austin, TX, USA
116. Interdisciplinary Life Sciences Graduate Programs (ILSGP), and Oden Institute for Computational Engineering and Sciences (ICES), The University of Texas at Austin, Austin, TX, USA
117. Department of Genetics, University of Pennsylvania, Philadelphia, PA., USA
118. Department of Electrical & Computer Engineering, University of Washington, Seattle, WA., USA
119. Department of Neuroscience, University of Pennsylvania, Philadelphia, PA., USA
120. Department of Medicine, University of Pennsylvania, Philadelphia, PA., USA
121. Department of Cancer Biology, University of Pennsylvania, Philadelphia, PA., USA
122. Developmental Biology Program, Sloan Kettering Institute, New York, NY, USA
123. Center for Stem Cell Biology, Sloan Kettering Institute for Cancer Research, New York, NY 10065, USA
124. Department of Biomedical Engineering and McKusick-Nathans Department of Genetic Medicine, Johns Hopkins University; Baltimore, MD 21218, USA
125. Department of Biochemistry, Albert Einstein College of Medicine, Bronx, NY 10461, USA
126. Weill Cornell Graduate School of Medical Sciences, Weill Cornell Medicine, 1300 York Avenue, New York, NY 10065, USA
127. Department of Pediatrics, University of California, San Diego, USA
128. Department of Medicine, University of California, San Diego, USA
129. Department of Cellular and Molecular Medicine, University of California, San Diego, CA, USA
130. Center for Epigenomics, University of California, San Diego

131. Bioinformatics and Systems Biology Program, University of California, San Diego, CA USA
132. Biomedical Sciences Program, University of California, San Diego, CA USA
133. Department of Human Genetics, University of California Los Angeles, Los Angeles, CA USA
134. Department of Biological Chemistry, David Geffen School of Medicine, University of California Los Angeles, Los Angeles, CA, USA
135. Department of Neurology, University of California Los Angeles, Los Angeles, CA, USA
136. Department of Pathology and Laboratory Medicine, University of California Los Angeles, Los Angeles, CA, USA
137. Department of Computational Medicine, University of California Los Angeles, Los Angeles, CA, USA
138. Department of Biostatistics, University of California Los Angeles, Los Angeles, CA, USA
139. Department of Genetics, Washington University, St. Louis, MO., USA
140. McDonnell Genome Institute, Washington University School of Medicine, Saint Louis, MO, USA
141. Sloan Kettering Institute, Memorial Sloan Kettering Cancer Center, New York, NY, USA
142. Center for Genetic Epidemiology, Department of Population and Public Health Sciences, Keck School of Medicine, University of Southern California, CA, USA
143. Department of Quantitative and Computational Biology, University of Southern California, CA, USA
144. Norris Comprehensive Cancer Center, Keck School of Medicine, University of Southern California, CA, USA
145. Institute of Computational Biology, Helmholtz Zentrum Munich, Neuherberg, Germany
146. Department of Computer Science, School of Computation, Information and Technology, Technical University Munich, Munich, Germany
147. Munich Heart Alliance, DZHK (German Center for Cardiovascular Research), Munich, Germany
148. Lawrence Berkeley National Laboratory, Berkeley, CA 94720
149. New York Genome Center, New York, NY USA
150. Department of Biology, New York University, New York, NY USA
151. DOE Joint Genome Institute, Berkeley, CA USA
152. Gladstone Institutes, San Francisco, CA USA
153. University of California, San Francisco, CA USA
154. Chan Zuckerberg Biohub - San Francisco, San Francisco, CA USA
155. St. Anna Children's Cancer Research Institute (CCRI), Vienna, Austria
156. CeMM Research Center for Molecular Medicine of the Austrian Academy of Sciences, Vienna, Austria
157. Livestrong Cancer Institutes, Department of Oncology, and Department of Biomedical Engineering, The University of Texas at Austin, Austin, TX, USA
158. Interdisciplinary Life Sciences Graduate Programs (ILSGP), and Oden Institute for Computational Engineering and Sciences (ICES), The University of Texas at Austin, Austin, TX, USA
159. Division of Genomic Medicine, National Human Genome Research Institute (NHGRI), National Institutes of Health (NIH), Bethesda, MD, USA
160. Division of Genome Sciences, National Human Genome Research Institute (NHGRI), National Institutes of Health (NIH), Bethesda, MD, USA
161. Department of Applied Physics, Stanford University, Stanford, CA, USA 94305
162. Cancer Biology Program, Stanford University School of Medicine, Stanford, CA, USA 94305
163. Immunology Graduate Program and Department of Pathology, Stanford University School of Medicine, Stanford, CA, USA 94305
164. Department of Plant Biology, Carnegie Institution for Science, Stanford, CA 94305, USA
165. Computational Biology Department, School of Computer Science, Carnegie Mellon University, Pittsburgh, PA, USA
166. Program in Computational Biology & Bioinformatics, Yale University, New Haven, CT, USA
167. Department of Molecular Biophysics & Biochemistry, Yale University, New Haven, CT, USA
168. Department of Computer Science, Yale University, New Haven, CT, USA
169. Department of Statistics & Data Science, Yale University, New Haven, CT, USA
170. Department of Biomedical Informatics & Data Science, Yale University, New Haven, CT, USA
171. Howard Hughes Medical Institute, Seattle, WA, USA

172. Systems Immunology, Benaroya Research Institute, Seattle, WA, USA
173. Allen Discovery Center for Cell Lineage Tracing, Seattle, WA, USA
174. mRNA Center of Excellence, Sanofi Pasteur Inc., Waltham, MA, USA
175. Department of Biochemistry and Molecular Genetics, Feinberg School of Medicine Northwestern University, Chicago, IL, USA
176. Robert H. Lurie Comprehensive Cancer Center of Northwestern University, Chicago, IL, USA
177. Center for Genetic Medicine, Department of Pharmacology, Northwestern University, Chicago, IL, USA
178. Center for Genetic Medicine and Department of Cell and Developmental Biology, Feinberg School of Medicine, Northwestern University, Chicago, IL, USA
179. Department of Preventive Medicine, Feinberg School of Medicine, Northwestern University
180. Department of Pediatrics, Washington University, St. Louis, MO., USA
181. Department of Medicine, Washington University, St. Louis, MO., USA
182. Department of Developmental Biology, Washington University, St. Louis, MO., USA
183. Department of Pathology and Immunology, Washington University, St. Louis, MO., USA
184. TERRA Teaching and Research Centre, University of Liège, Gembloux, Belgium
185. Computational Biology and Bioinformatics, Université Libre de Bruxelles, Brussels, Belgium
186. UC Santa Cruz Genomics Institute, University of California Santa Cruz, Santa Cruz, CA, USA
187. Gladstone Institute of Data Science and Biotechnology, Gladstone Institutes, San Francisco, CA, USA
188. Department of Developmental and Cell Biology, UC Irvine, Irvine, CA., USA
189. Center for Complex Biological Systems, UC Irvine, Irvine, CA., USA
190. Division of Biology and Biological Engineering, California Institute of Technology, Pasadena, CA USA
191. Richard N. Merkin Institute for Translational Research, California Institute of Technology, Pasadena, CA., USA
192. Division of Chemistry and Chemical Engineering, California Institute of Technology, Pasadena, CA., USA
193. Department of Neurobiology and Behavior, UC Irvine, Irvine, CA., USA
194. Department of Computing and Mathematical Sciences, California Institute of Technology, Pasadena, CA., USA
195. Department of Pharmaceutical Sciences, UC Irvine, Irvine, CA., USA
196. Department of Computational Medicine and Bioinformatics, University of Michigan, Ann Arbor, MI, USA
197. Department of Human Genetics, University of Michigan, Ann Arbor, MI, USA
198. Department of Biostatistics, School of Public Health, University of Michigan, Ann Arbor, MI, USA
199. Computational and Systems Biology Program, Memorial Sloan Kettering Cancer Center, New York, NY, USA
200. Howard Hughes Medical Institute and Immunology Program at Sloan Kettering Institute, New York, NY, USA
201. Ludwig Center for Cancer Immunotherapy, Memorial Sloan Kettering Cancer Center, New York, NY, USA
202. Division of Rheumatology, Department of Medicine, Hospital for Special Surgery, New York, NY, USA
203. Weill Cornell Medical College and Graduate School, New York, NY, USA
204. Department of Biostatistics, University of North Carolina at Chapel Hill, Chapel Hill, NC, USA
205. Department of Pharmacology, University of North Carolina at Chapel Hill, Chapel Hill, NC, USA
206. Department of Environmental Health Sciences, Mailman School of Public Health Columbia University, New York, NY, USA
207. Masters of Physician Assistant Studies Program, Colorado Mesa University, Grand Junction CO, USA
208. Department of Immunology, University of Washington, Seattle, WA, USA
209. Cincinnati Children's Hospital, Cincinnati OH, USA
210. Department of Genetics and Development, Institute for Cancer Genetics, Herbert Irving Comprehensive Cancer Center, Columbia University Irving Medical Center, New York, NY, USA
211. Department of Neuroscience, University of Texas Southwestern Medical Center, Dallas, TX, USA

212. Department of Psychiatry, University of Texas Southwestern Medical Center, Dallas, TX, USA
213. Center for the Genetics of Host Defense, University of Texas Southwestern Medical Center, Dallas, TX, USA
214. Peter O'Donnell Jr. Brain Institute, University of Texas Southwestern Medical Center, Dallas, TX, USA
215. Altos Labs Inc., 1300 Island Drive, Redwood City, CA, USA
216. Department of Neurology, Washington University, St. Louis, MO., USA
217. Department of Pathology, Massachusetts General Hospital, Boston, MA, USA
218. PhD Program in Biological and Biomedical Sciences, Harvard University, Boston, MA, USA
219. Departments of Psychiatry and Genetics, Division of Molecular Psychiatry, Department of Genetics, Wu Tsai Institute, Yale University School of Medicine, New Haven, CT, USA
220. MIT Computer Science and Artificial Intelligence Laboratory, Massachusetts Institute of Technology, Cambridge, MA, USA
221. Paul G. Allen School of Computer Science and Engineering, University of Washington, Seattle, WA, USA
222. Canadian Institute for Advanced Research, Toronto, ON, Canada
223. Department of Genetics, Yale School of Medicine, New Haven, CT, USA

## Author Contributions

Jesse M. Engreitz, Heather A. Lawson, and Harinder Singh co-led the Writing Group. Jesse M. Engreitz, Heather A. Lawson, Harinder Singh, Lea Starita, Gary C. Hon, Hannah Carter, Nidhi Sahni, Timothy E. Reddy, Xihong Lin, Yun Li, Nikhil Munshi, Maria Chahrour, Benjamin Hitz, and Ali Mortazavi wrote initial text based on input from PIs, the Writing Group, and Working Group and Focus Group Co-Chairs. Jesse M. Engreitz, Alan Boyle, and Jayoung Ryu developed figures. All authors contributed to developing the vision and goals of the IGVF Consortium, outlining the project, and editing the manuscript. The role of the NHGRI Program Management in the preparation of this paper was limited to coordination and scientific management of the IGVF Consortium.

## Acknowledgements

This work was supported by the NIH NHGRI IGVF Program (UM1HG011966, UM1HG011969, UM1HG011972, UM1HG011989, UM1HG011996, UM1HG012003, UM1HG012010, UM1HG012053, UM1HG011986, UM1HG012076, UM1HG012077, U01HG011952, U01HG011967, U01HG012009, U01HG012022, U01HG012039, U01HG012064, U01HG012069, U01HG012041, U01HG012047, U01HG012051, U01HG012059, U01HG012079, U01HG012103, U24HG012012, U24HG012070), NIH NCI (R01CA197774), and the Novo Nordisk Foundation (NNF21SA0072102). Figures were illustrated by SciStories LLC. We thank members of the IGVF External Consultants Panel (Guillaume Bourque, Prashant Mali, Judy Cho, Barbara Engelhardt, and Olga Troyanskaya) for critical feedback on the manuscript.